\documentclass{iscram}

\usepackage[utf8]{inputenc}
\usepackage{lipsum}
\usepackage{tikz}
\usepackage{fancyvrb}
\usepackage{listings}
\usepackage{enumitem}
\usepackage{tcolorbox}
\usepackage{multicol}
\usepackage{siunitx}
\usepackage{graphics}
\usepackage{multirow}
\usepackage{booktabs}
\usepackage{graphicx}
\usepackage{float}
\usepackage{caption}
\usepackage{subcaption}
\usepackage{chato-notes}
\graphicspath{{./figures/}}

\lstdefinestyle{common}{
  xleftmargin=.5em,
  xrightmargin=.5em,
  frame=single,framesep=.5em,framerule=0pt,
  fancyvrb=true,
  basicstyle=\ttfamily,
  keywordstyle=\color{cyan!50!blue!75!black}\bfseries,
  commentstyle=\color{red!50!black}\itshape,
  stringstyle=\ttfamily\color{green!50!black},
  numbers=none,
  showspaces=false,
  showstringspaces=false,
  fontadjust=true,
  keepspaces=true,
  flexiblecolumns=true,
  emphstyle=\color{red},
}
\lstdefinestyle{TeX}{
  style=common,
  backgroundcolor=\color{blue!5},
  aboveskip=5pt,
  belowskip=5pt,
  language=[LaTeX]TeX,
  moretexcs={
    abstract, addbibresource, iscramset, keywords, mainmatter,
    maketitle, printbibliography, subsection, subsubsection, url,
    urldef, href, includegraphics, ldots, parencite, citeauthor,
    citeyear, citetitle, midrule, toprule, bottomrule
  },
  fancyvrb=true,
}
\lstdefinestyle{console}{
  style=common,
  backgroundcolor=\color{gray!10},
  aboveskip=5pt,
  belowskip=5pt,
}

\newlist{options}{description}{1}
\setlist[options]{%
  beginpenalty=10000,%
  itemsep=.5\parskip plus .3\parskip minus .2\parskip,
  parsep=.5\parskip plus .3\parskip minus .2\parskip,
  topsep=.5\parskip plus .3\parskip minus .2\parskip,
  partopsep=.5\parskip plus .3\parskip minus .2\parskip,
  style=nextline,labelindent=1em,%
  font=\normalfont\ttfamily}

\colorlet{macro color}{cyan!50!blue!75!black}
\colorlet{option color}{red!50!black}
\colorlet{generic color}{green!40!black}

\newcommand\generic[1]{\textcolor{generic color}{\ttfamily\itshape\makebox{<#1>}}}
\newtcolorbox{pseudoTeX}{colback=blue!5,colframe=blue!5,before=\nobreak}
\let\LaTeXorig\LaTeX
\renewcommand\LaTeX{\bgroup\fontfamily{lmr}\selectfont\upshape\LaTeXorig\egroup}

\urldef{\sitewebgind}\url{http://gind.mines-albi.fr}
\urldef{\sitewebminesalbi}\url{http://www.mines-albi.fr}
\urldef{\gmailpaulgaborit}\url{paul.gaborit@gmail.com}
\urldef{\jdmail}\url{j.doe@example.com}
\urldef{\sitex}\url{www.example.com}

\iscramset{
  title={
    Uneven Coverage of Natural Disasters in Wikipedia: the~Case~of~Floods
  },
  short title={Uneven coverage of natural disasters in Wikipedia},
   author={
    short name={Lorini},
    full name=Valerio Lorini,
    affiliation={
      European Commission, Joint Research Centre (JRC), Ispra, Italy \\
      Universitat Pompeu Fabra, Barcelona, Spain \\
      {valerio.lorini@ec.europa.eu}
    },
  },
  author={
    full name=Javier Rando,
    affiliation={
      Universitat Pompeu Fabra, Barcelona, Spain \\
      javier.rando01@estudiant.upf.edu
    },
  },
  author={
    full name=Diego Saez-Trumper,
    affiliation={
      Wikimedia \\
      diego@wikimedia.org
    },
  },
  author={
    full name=Carlos Castillo,
    affiliation={
      Universitat Pompeu Fabra, Barcelona, Spain \\
      chato@acm.org
    },
  },
  footer/line 1={\emph{CoRe Paper} -- \emph{Social Media for Disaster Response and Resilience}},
}

\usepackage{tikzpagenodes}
\usetikzlibrary{fit}

\newcommand{\spara}[1]{\smallskip\noindent{\bf #1}}

\begin{document}

\maketitle

\makeatletter
\makeatother

\abstract{
The usage of non-authoritative data for disaster management presents the opportunity of accessing timely information that might not be available through other means, as well as the challenge of dealing with several layers of biases.
Wikipedia, a collaboratively-produced encyclopedia, includes in-depth information about many natural and human-made disasters, and its editors are particularly good at adding information in real-time as a crisis unfolds. In this study, we focus on the English version of Wikipedia, that is by far the most comprehensive version of this encyclopedia.
Wikipedia tends to have good coverage of disasters, particularly those having a large number of fatalities.
However, we also show that a tendency to cover events in wealthy countries and not cover events in poorer ones %
permeates Wikipedia as a source for disaster-related information.
By performing careful automatic content analysis at a large scale, we show how the coverage of floods in Wikipedia is skewed towards rich, English-speaking countries, in particular the US and Canada.
We also note how coverage of floods in countries with the lowest income, as well as countries in South America, is substantially lower than the coverage of floods in middle-income countries.
These results have implications for systems using Wikipedia or similar collaborative media platforms as an information source for detecting emergencies or for gathering valuable information for disaster response.
}


\section{Introduction}
\label{sec:introduction}

During the past decade, water-related disasters happened more frequently and their impacts have been more severe, hindering sustainable development by causing socio-economic disruption in many countries.
Altogether, water-related disasters, such as floods, droughts, storm surges, cyclones, convective storms and tsunamis, account for 90\% of all disasters in terms of the number of people affected and among them, 50\% are flood events \parencite{HELP_2019}.

Unaddressed vulnerabilities, rising population, intertwined natural events, continue to be the main critical factors for catastrophic loss of life, disrupting livelihoods and fueling new displacement.
A previous analysis estimated that people in the least developed countries are, on average, six times more likely to be injured, lose their home, be displaced or evacuated, or require emergency assistance, than those in high-income countries \parencite{UNDRR_2019}.

Death tolls and economic losses from natural hazards are expected to rise in many parts of the world.
Several possible future scenarios show that flood impacts will increase. The more rapid rise of fatalities will be in countries with high human vulnerability, which is a combination of high exposure, low flood protection, and high mortality rates.
Countries with higher income levels show lower human vulnerability and the high number of people exposed translates into lower mortality compared to developing countries \parencite{dottori_increased_2018}.
An analysis of vulnerability at a global scale, integrating population and economic dynamics with one of the most comprehensive natural disaster loss databases, show that there is still a considerable climate hazard vulnerability gap between poorer and wealthier countries  \parencite{Formetta2019}.

Wikipedia, founded in 2001, has come a long way, becoming over the years one of the primary sources of encyclopedic information worldwide. In fact, during 2018 alone, the English Wikipedia had over 108 billions article views. It accounts for approximately 45\% of all page views on Wikimedia projects in this period (237B)\footnote{\url{https://stats.wikimedia.org/v2}}.

One of Wikimedia's goal is sharing knowledge, and an extensive international base of editors is a crucial element in providing information in several languages.
Even if a useful, ethical code for Wikipedians can guide editing towards styles of practice that best support the Wikipedia mission\footnote{\url{https://en.wikipedia.org/wiki/Wikipedia:Ethical_Code_for_Wikipedians}}, when editors mix personal interests with the goals of the Wikipedia community as a whole, they make choices that can affect the articles they create and edit \parencite{Hube:2017:BW:3041021.3053375}.
Although collaborative editing fulfils the objective of sharing information, it can introduce biases that are apparent when Wikipedia is used as a reference data set for a specific topic, such as natural science research.

In recent years, researchers have placed much effort into studying how to extract meaningful information for crisis management from social media and collaborative sources  \parencite{lorini2019integrating,imran2016twitter,olteanu_what_2015}, but biases in these sources are rarely evaluated. %

In a seminal paper, Galtung and Ruge (\citeyear{Galtung_1965}) showed that pieces of news from ``elite'' nations were more likely to be covered in foreign news reports. We find evidence of the same for the coverage of floods in the English Wikipedia, noting that floods in the wealthiest countries, particularly floods in the US, are more likely to appear in Wikipedia than floods in the poorest countries.

We think that Wikipedia is a valuable source of free data, and it could be beneficial to researchers in the Disaster Risk Reduction field if biases are identified, measured, and mitigated.
In this article, we estimate the coverage of floods in Wikipedia along many variables, including Gross Domestic Product (GDP), Gross National Income (GNI), geographic location, number of English speakers,  fatalities, and various indices describing the level of vulnerability of a country.

Addressing flaws and exposing biases can help the research community to think about possible countermeasures that can lead to a set of best practices for Wikipedia or for publishing research leveraging Wikipedia data.
Our main contributions are:
\begin{itemize}
\item We establish a validated reference set of events tracked by several independent organizations, with support from hydrologists.
Some organizations collect data about floods for different purposes, from insurance to sustainable development goals set by the United Nations (UN). Their effort is to collect floods data on a global scale. We compare and collate the different data sources.
\item We match verified events with Wikipedia entries.
We analyze three methodologies for matching verified events with Wikipedia's  text in terms of location and temporal references.
In our work, a particular effort has been made to geo-locate Wikipedia entry candidates since we wanted to identify news reporting information about an event and to exclude generic collections of unspecified events.
\end{itemize}

The remainder of this paper is organized as follows:
the next section presents related work;
then, the third section describes the methods for establishing verified ground truth information, for matching Wikipedia data with verified events and how to geo-locate them
Finally, we present experimental results, including a case study, followed by our conclusions and future work.

\section{Related Work}
\label{sec:related}

Wikipedia has been used as a data source to study \emph{sustainable development} and for \emph{Disaster Risk Reduction.}
For instance, it has been recently used as a source of data to estimate  indicators at very high spatial resolution leveraging recent advances in Natural Language Processing (NLP) by extracting information from free-text articles \parencite{Sheehan:2019:PED:3292500.3330784}.
In their work, the spatial distribution of the articles and meta-data extracted from its text, combined with other data such as night-light satellite images, are used to improve the  prediction of socio-economic indicators of poverty, as measured by ground truth survey data collected by the World Bank.

In previous work, researchers used Wikipedia for detecting and monitoring natural disasters \parencite{44015} leveraging interlinks between versions of the same article in different languages and inbound/outbound redirects to other similar articles.
The methodology proposed in their paper consists in creating and maintaining a list of articles related to natural disasters, scanning Wikipedia entries and subsequently checking if edits happen on an article in the list, assuming a new event reported  would impact the monitoring-list.

Considering that Wikipedia is being used as a source for data analysis, our work aims at identifying potential biases in Wikipedia coverage of natural events, specifically floods.

Wikipedia exhibits a substantial amount of self-focus, in the sense that editors in each language-specific Wikipedia tend to write about topics that are of interest to their community and not others \parencite{hecht2009measuring}.
A country-based analysis of Wikipedia shows that geotagged articles (i.e., articles referring to specific locations) concentrate in only a few countries, and this concentration can be explained in no small extent with variables such as population, number of broadband Internet connections, and number of edits emanating from each country \parencite{graham2014uneven}.

A comparison of Wikipedia with the Global Terrorism Database\footnote{\url{https://www.start.umd.edu/data-tools/global-terrorism-database-gtd}} in 2015 shows that Wikipedia covered about 78\% of attacks and almost all of the terrorism-related deaths in Western countries, but only 28\% of those in other countries \parencite{rossling2019factfullness}.

Also, Wikipedia suffers from a cultural gap that favours entries written in English and especially, those referring to the United States of America (USA) which are the longest and best-referenced ones \parencite{doi:10.1002/asi.21577}.
Tobler's law for geography claims that similarity decreases with distance \parencite{tobler1970computer}. According to this law, those events happening close to English speaking countries should be considered more familiar to Wikipedia editors and therefore, better covered than those happening in distant places. There are also urban/rural biases, with Wikipedia coverage of rural areas being systematically inferior in quality \parencite{johnson2016not}.

Becoming a source on current news events was not part of the original mission of Wikipedia, but currently, the most visited and edited articles are about current events \parencite{keegan2012staying}. Wikipedia has transitioned into a source that incorporates significant news work \parencite{keegan2013history}.

When it comes to history, Wikipedia narratives are biased towards recent events and those happening in Western Europe \parencite{samoilenko2017analysing,Samoilenko2017History}. Partially because of this, there is an explicit Wikipedia policy against ``recentism''\footnote{\url{https://en.wikipedia.org/wiki/Wikipedia:Recentism}}.

Regarding coverage of natural disasters, a study on the Tōhoku catastrophes showed that activity on Wikipedia concentrated on the day of the earthquake, but there was intense editing activity for several days \parencite{keegan2011hot}. A similar pattern of intense activity close to the events was observed in the 2011 Arab spring \parencite{ferron2011collective}. Most of these event-centric articles are written as the event unfolds \parencite{moats2019following} and indeed, spikes in editing activity can be used for detecting new crisis events \parencite{steiner2015disaster}.

Our work is focused on natural disasters at a global scale for events happening over more than three years. Therefore, our experiments widen the previous analysis of biases, including a set of socio-economic risk indicators concerning natural hazards.

\section{Methods}
\label{sec:methods}

\subsection{Ground Truth Generation}
This work aims to analyze the coverage of floods in Wikipedia.
To carry out our experiments, we established a consistent and reliable ground truth about flooding events globally. We developed methodologies for extracting candidates from Wikipedia articles and defined metrics for evaluating matches between candidates and ground truth information.

Development and relief agencies have long recognized the crucial role played by data and information from previous events in mitigating the impacts of disasters on vulnerable populations. Due to the complexity of collecting reliable information, there is still no international consensus regarding best practices for defining critical aspects of an event such as starting date, duration or number of fatalities.

To carry out our experiments, we selected data source which included  validated information from international relief agencies or local governments worldwide to cover all the events that could have been detected on the social networks so Precision and Recall could be computed against a complete validated dataset.
That is why we consider three of the most comprehensive databases documenting floods that are commonly used by the hydrology science for reference\parencite{wirtz_need_2014}:

\begin{itemize}[noitemsep,topsep=1pt]
    \item \textbf{Floodlist}\footnote{\url{https://floodlist.com/}},
    funded by the EU Space program Copernicus \footnote{\url{https://www.copernicus.eu/en}} program, it reports on all the major flood events from around the world.
    Floodlist includes articles on flood-related issues such as warning systems, mitigation and control, flood recovery, flood damage repair and restoration, as well as flood insurance. The reports and articles also include information about the extraordinary humanitarian, aid and relief efforts made in the aftermath of many flood disasters.
    \item The Emergency Events Database (\textbf{EM-DAT})\footnote{\url{https://www.emdat.be}} contains information from various sources, including UN agencies, non-governmental organizations, insurance companies, research institutes and press agencies. Data from UN agencies, governments, and the International Federation of Red Cross and Red Crescent Societies have priority. This choice is not only a reflection of the quality or value of the data, but it also reflects the fact that most reporting sources do not cover all disasters or have political limitations that could affect the figures. The entries are reviewed continuously for inconsistencies, redundancy, and incompleteness. CRED consolidates and updates data daily.
    \item The Dartmouth Flood Observatory (\textbf{DFO})\footnote{\url{ https://http://floodobservatory.colorado.edu/}},
    based at the University of Colorado, maintains the Global Active Archive of Large Flood Events derived from news, governmental, instrumental, and remote sensing sources. The archive is ``active'' because current events are added immediately.
    Each entry in the archive and related ``area affected'' map outline represents a discrete flood event. The listing is comprehensive and global in scope.
\end{itemize}

We also looked at other reliable sources such as the Copernicus Emergency Manamagent Services (CEMS) based on requests for satellite images acquisition for emergency response and risk and recovery maps.
We found that official requests were issued only when the national authorities could not cope with the disaster by their meaning, resulting in only one-tenth of events recorded by the other sources, most of which were redundant.

As shown in Table~\ref{tab:totflood}, none of the three selected databases is complete, and some events recorded in one database are not in the others. Hence, we merge multiple databases into a single dataset. Our data begins on 2016-02-29 because this is the earliest date for which the three datasets contain information. %
\begin{table}[htb]
\centering
\begin{tabular}{@{}llllll@{}}
\toprule
 Year & Total floods & Floodlist  & EM-DAT & DFO \\ \midrule
 2016 & 261 & 191 & 169 & 99 \\
 2017 & 322 & 220 & 215 & 117 \\
 2018 & 394 & 306 & 191 & 157 \\
 2019 & 125 & 96 & 74 & 42 \\
\midrule
Total & 1102 & 813 & 415 & 649 \\ \bottomrule
\end{tabular}
\caption{Number of events recorded per year in chosen data sources between 2016-02-29 and 2019-05-20}
\label{tab:totflood}
\end{table}

\spara{Criteria and Definitions of Events}
It is essential to assess criteria for event recording and limitations of the several data sources before homogeneously merging their data.
\begin{itemize}[noitemsep,topsep=1pt]
    \item %
    Floodlist includes articles on flood-related issues such as warning systems, mitigation and control, flood recovery, flood damage repair and restoration, as well as flood insurance. A small team of dedicated people record news and information on the latest flood events from around the world.
    We decided to leave out news items with ``landslides'' as the only tag, while we ingested all the other news items as we think they were mostly relevant to floods.
    \item %
    In EM-DAT, for a disaster to be entered into the database at least one of the following criteria must be fulfilled:
    (i) ten (10) or more people reported killed,
    (ii) one hundred (100) or more people reported affected,
    (iii) a declaration of a state of emergency, and/or
    (iv) a call for international assistance.
    EM-DAT provides geographical, temporal, human and economic information on disasters at the country-aggregated level. When the same disaster affects several countries, EM-DAT enters several country-level disasters into the database. Each of these country-level disasters will have the same identifier.
    From all the EM-DAT database, we consider only events labelled with ``flood'' or ``storm'' as primary disaster type.

    \item %
    DFO %
    derives from a wide variety of news and governmental sources. The quality and quantity of information available about a particular flood are not always in proportion to its actual magnitude, and the intensity of news coverage varies from nation to nation.
    DFO creates a record for any flood that appears to be ``large'', with, for example, significant damage to structures or agriculture, long (decades) reported intervals since the last similar event, and/or fatalities.
    Deaths and damage estimates for tropical storms are totals from all causes, but tropical storms without significant river flooding are not included.
    No filter is applied to information as we assumed all the news items were relevant to floods.
\end{itemize}

Since our ground truth information's main purpose is to support the analysis of the coverage of events in Wikipedia, we opted for a rather inclusive definition of flood and included events associated to heavy rainfall, which is the first driver of an overflow of water in river channels but also in coastal and urban areas.
The merged database contains information from different sources, trying to avoid duplicates.
We aggregated events at the national level, and when an event affected more than one country, we insert a record for each country with the same dates.
For those events for which a data source did not indicate the end date, we  assumed it was three days after the starting date of the event. We choose three because it is the median value of the duration of the floods in our dataset.
An example record contains the following information:
\begin{itemize}[noitemsep,topsep=1pt]
\item start\_date: year, month, day
\item end\_date: year, month, day
\item country: name of the country
\item affected: string from the source about population affected
\item fatalities: number of deaths associated with the event
\item location\_source1: location from location\_source1
\item location\_source2: location from location\_source2
\item location\_source3: location from location\_source3
\item identifier: list of id from sources
\item disaster\_type: i.e Storm, Flash Flood, Flood
\item in\_emdat: True/False
\item in\_dartmouth: True/False
\item in\_floodlist: True/False
\label{itm:record}
\end{itemize}

The resulting ground truth dataset contained 2295 floods. However, there were still many duplicate items needing to be consolidated.

\subsection{Event Consolidation}

As shown in the previous section, the criteria for describing an event vary according to the source.
Starting and ending dates are difficult to establish and there is no agreed methodology for reporting about duration and impacts of floods, not even within the Disaster Risk Reduction community.

For instance, a flood in Angola on March 2016 was reported by Floodlist as spanning the period 2016-03-05 to 2016-03-07, by EM-DAT as happening from 2016-03-01 to 2016-03-10, and by DFO as occurring from 2016-03-01 to 2016-03-10.

To avoid duplicates, when two or more events from the data sources overlapped in time and country, the earliest starting date was selected as starting date for the event and the latest ending date as the ending date.
This choice also means that two events happening at an overlapping time in two different locations of the same country will be considered as  duplicates and merged.
The aggregation was because the DFO dataset can locate an event only at the national level.
We normalized country names of each source to facilitate the merging process.

\begin{figure}[H]
\centering
  \includegraphics[width=.45\linewidth]{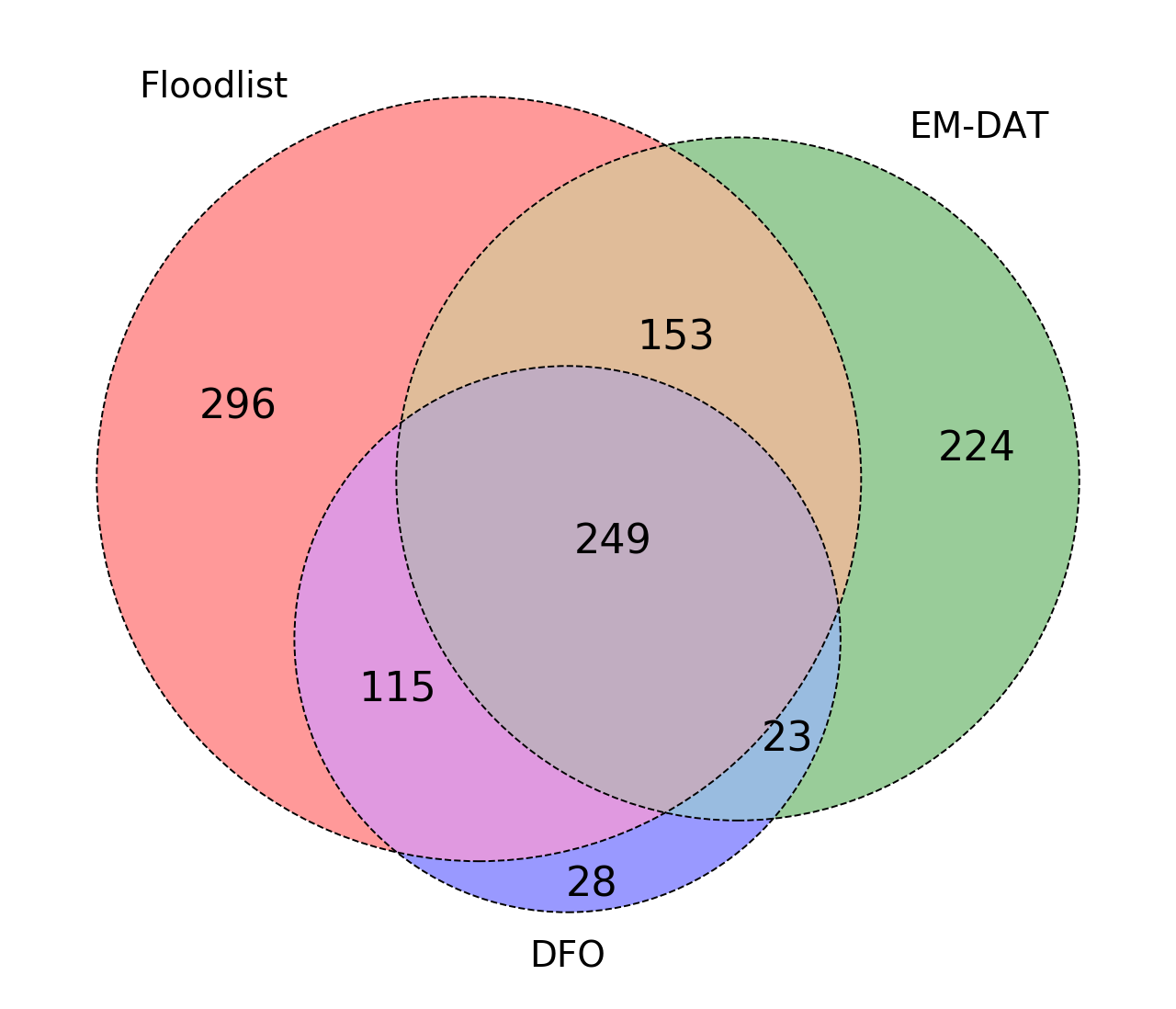}
  \caption{Venn diagram representing the intersection of floods coming from the data sources}
  \label{fig:venn}
\end{figure}

After the consolidation process, the dataset consisted of 1102 flood events.
Figure~\ref{fig:venn} shows a Venn diagram illustrating the intersection between and among our three data sources.
The intersecting areas are consistent and represent the majority of events, meaning that more than one source identified such floods.
We asked experts in the field of DRR to analyze a sample of twenty records that appeared only in one data source. They convened that:
\begin{itemize}[noitemsep,topsep=0pt]
  \item Unique records from Floodlist were mostly due to the inclusion of landslide associated with storms and episodes of heavy rain which were excluded from the other two data sources because not defined by their criteria as a flood.
  \item Unique records from EM-DAT were due to the inclusion of convective storms that lead to wind storms or sand storms which were excluded from the other two data sources because not associated with a flood.
  \item Unique records from DFO were mainly due to a country attribution different than the other sources in case of transborder events.
\end{itemize}
In light of the analysis of the data sources, we decided to conduct the experiments using the 458 events located in the intersecting areas, assuming that we can safely consider floods recorded by two or three data sources.

\begin{figure}[H]
\centering
   \begin{subfigure}{0.45\textwidth}
     \centering
     \includegraphics[width=\linewidth]{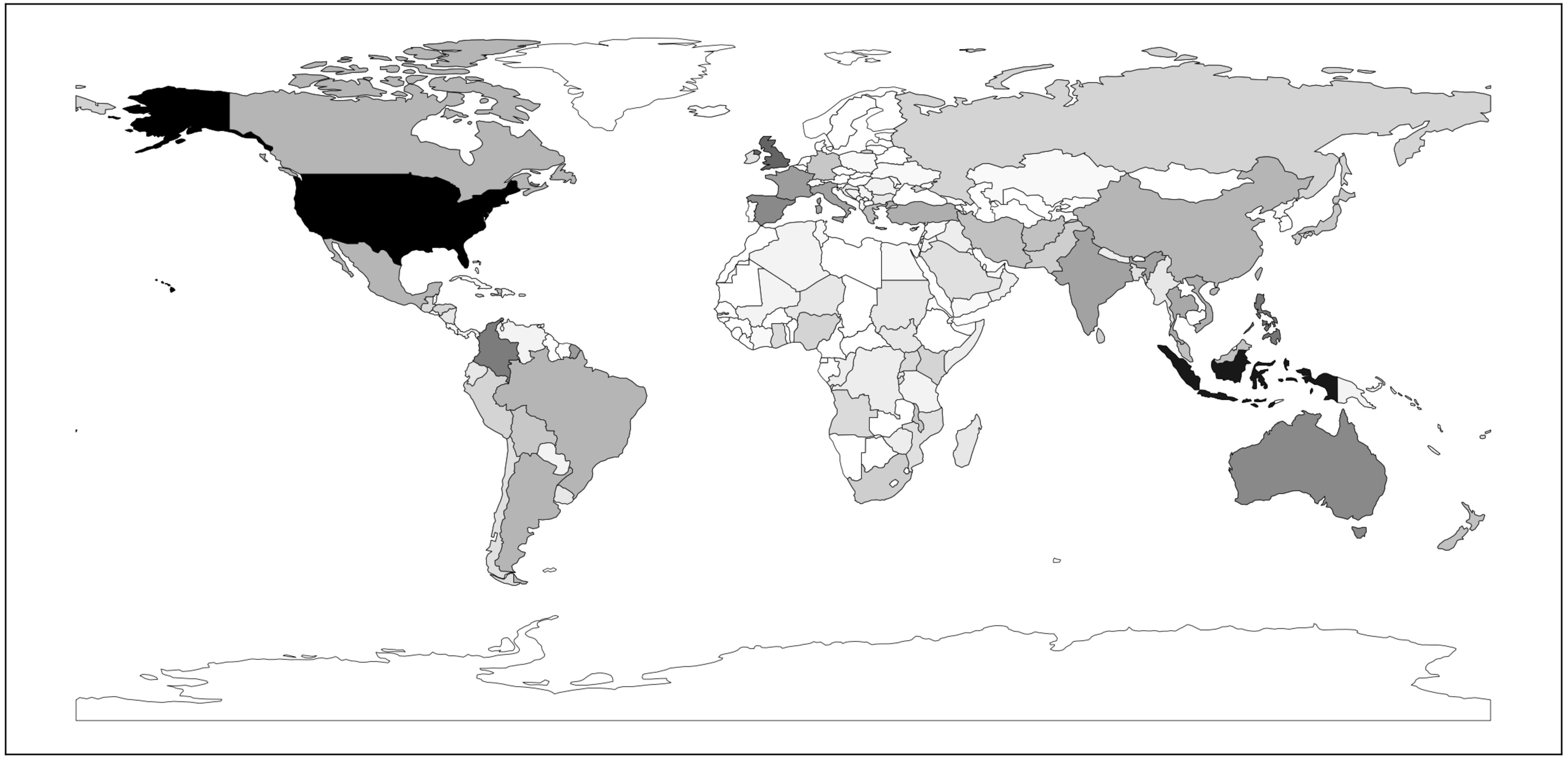}
     \caption{Floods per country - Floodlist}
     \label{fig:floodlistmap}
   \end{subfigure}
   \begin{subfigure}{0.45\textwidth}
     \centering
     \includegraphics[width=\linewidth]{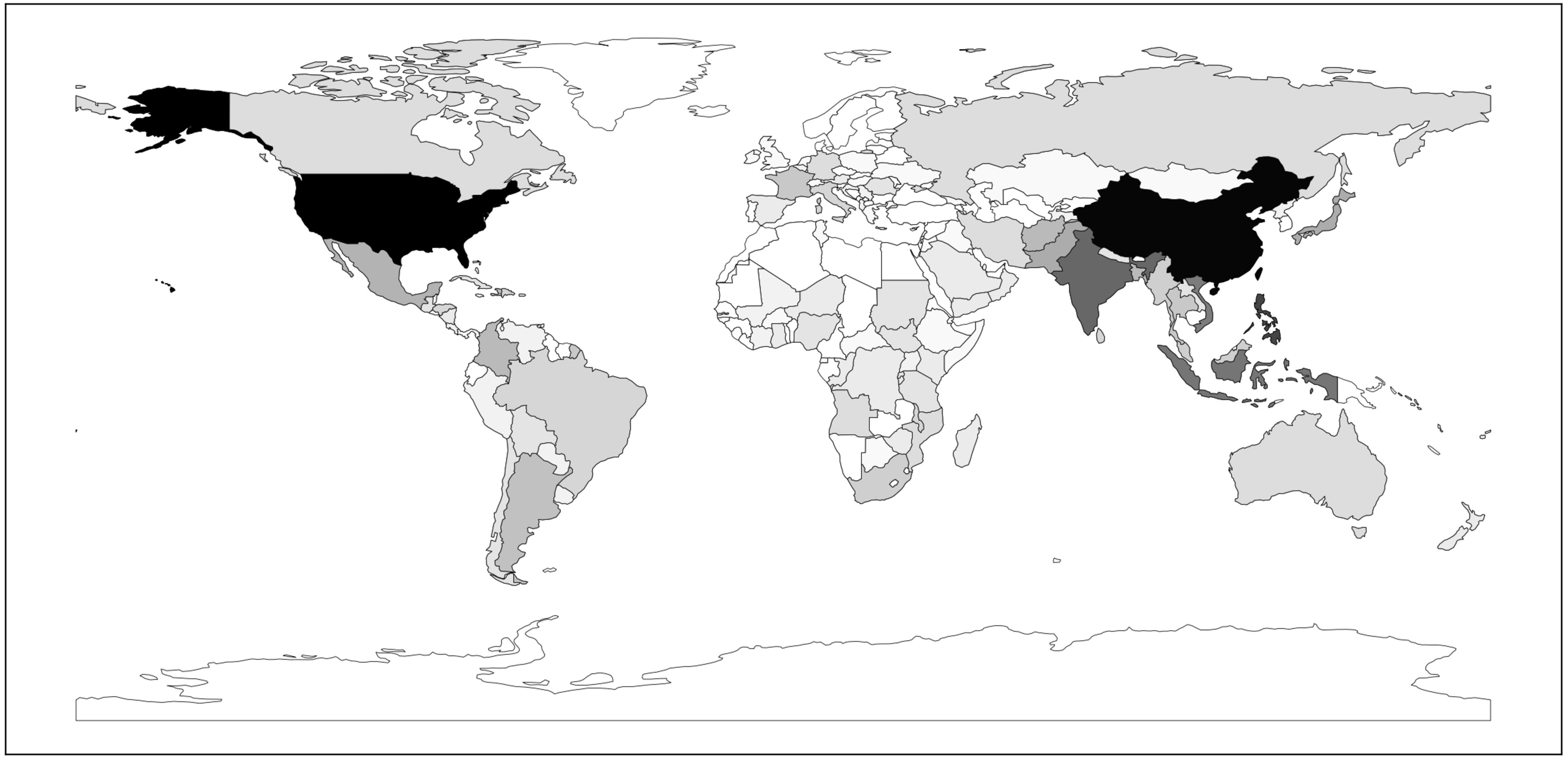}
     \caption{Floods per country - EM-DAT}
     \label{fig:emdatmap}
   \end{subfigure}
   \newline
   \begin{subfigure}{0.45\textwidth}
     \centering
     \includegraphics[width=\linewidth]{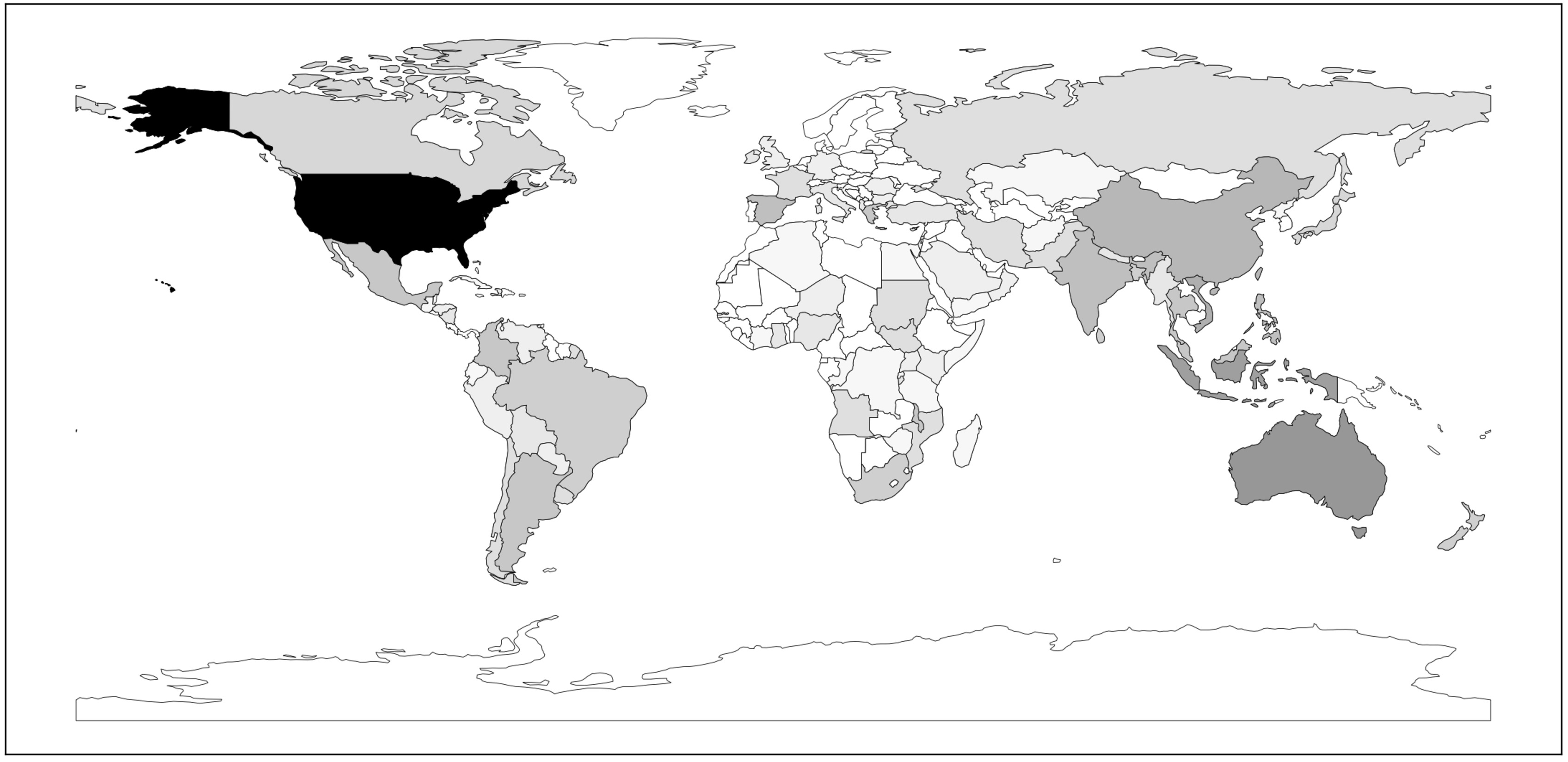}
     \caption{Floods per country -  DFO}
     \label{fig:dfomap}
   \end{subfigure}
   \begin{subfigure}{0.45\textwidth}
     \centering
     \includegraphics[width=\linewidth]{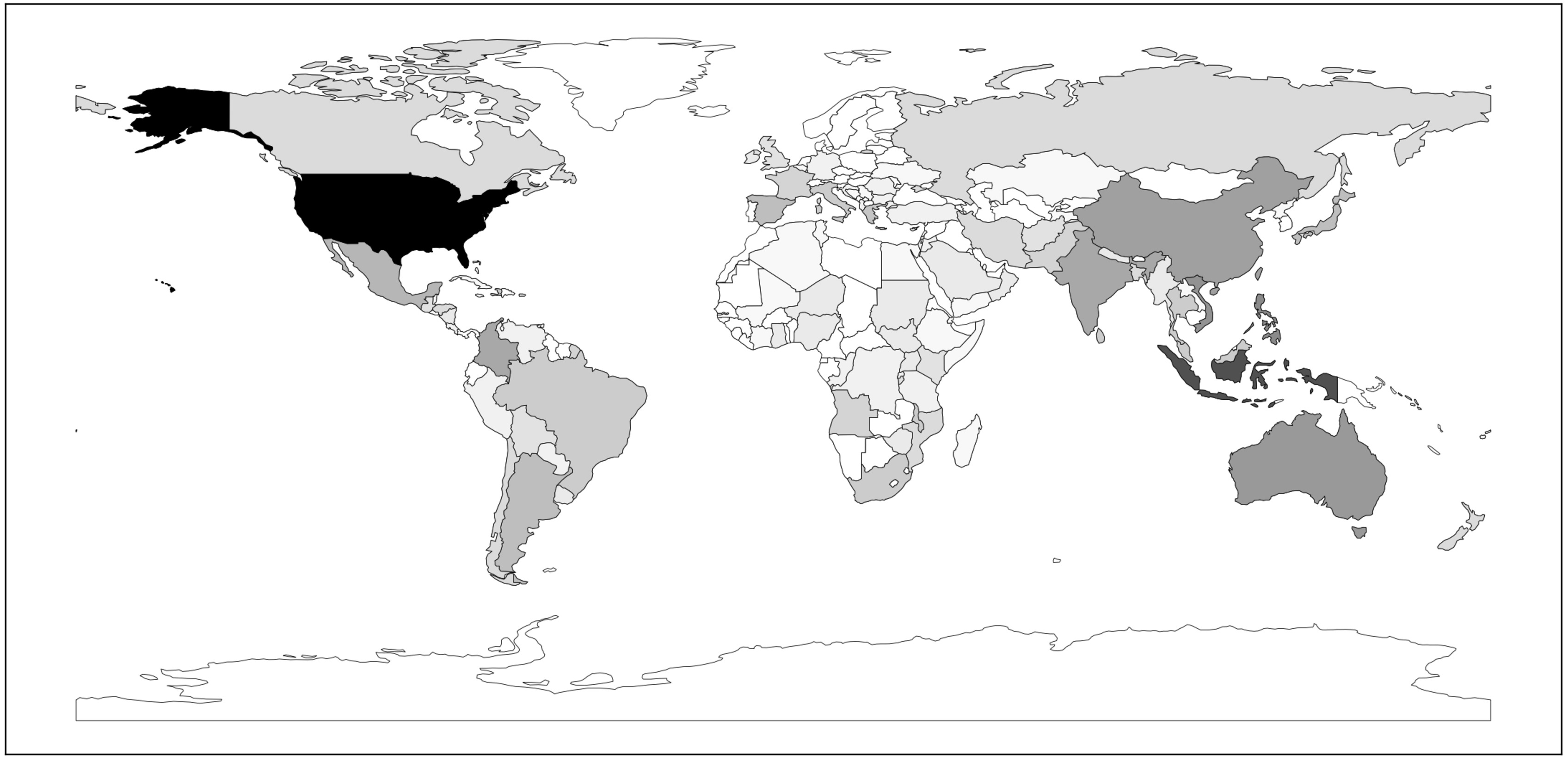}
     \caption{Floods per country - final merged ground truth }
     \label{fig:intersection}
   \end{subfigure}
   \newline
   \begin{subfigure}{0.90\textwidth}
     \centering
     \includegraphics[width=\linewidth]{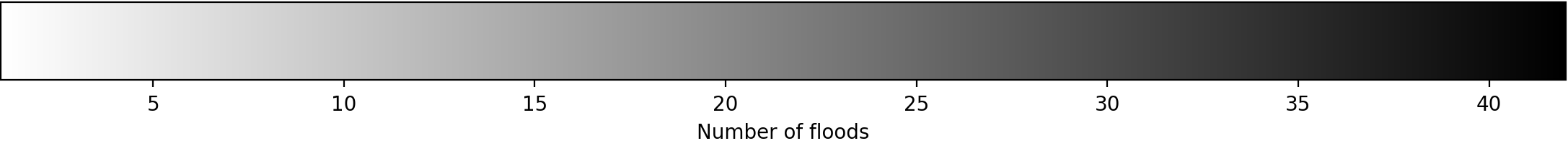}
   \end{subfigure}
   \caption{Floods per country (heat maps) of the three data sources for ground truth information and the final merged dataset}
   \label{fig:heatmaps}
\end{figure}

Figure \ref{fig:heatmaps} shows the geographical distribution of events recorded in each data source and the final merged result

\begin{figure}[H]
    \centering
    \includegraphics[width=.8\linewidth]{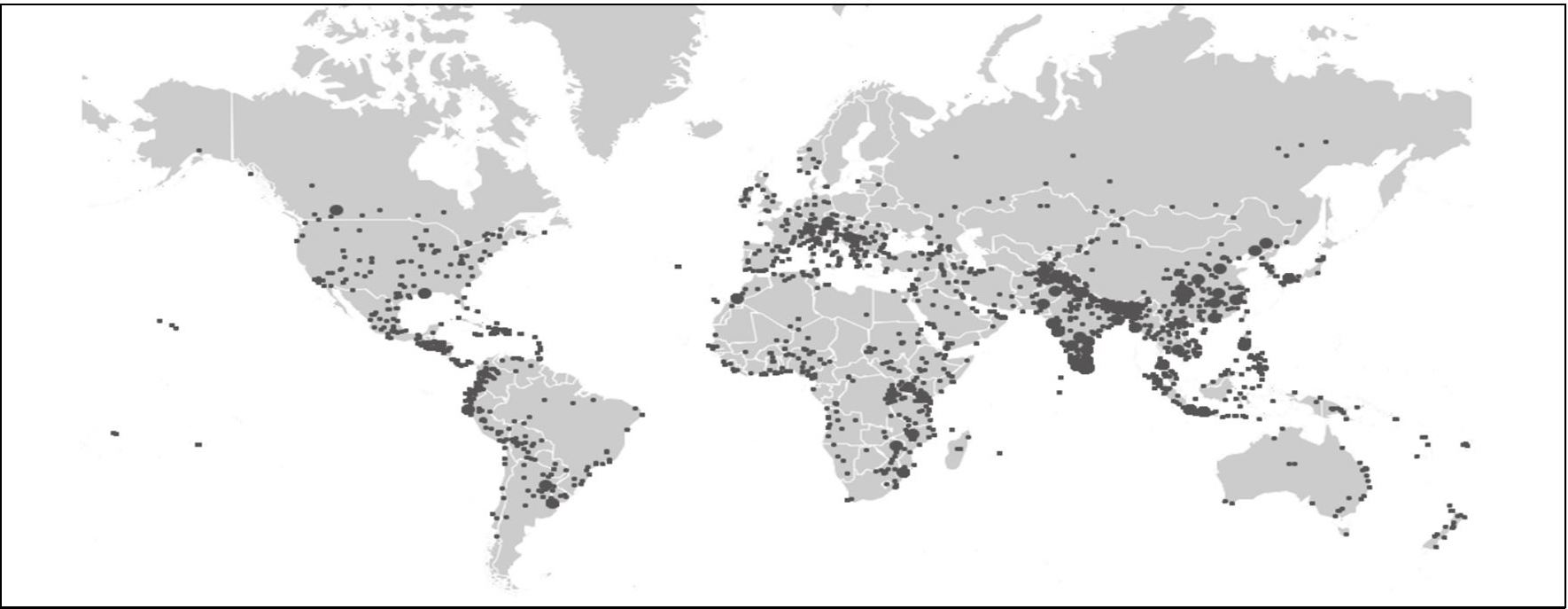}
    \caption{hydrological events listed in NatCatSERVICE for the period 2013-2018}
    \label{fig:munichre}
\end{figure}

To further evaluate potential biases in the distribution of events across data sources we compared the ground truth information with data publicly available from Munich RE’s NatCatSERVICE\footnote{\url{https://natcatservice.munichre.com/}}, one of the most comprehensive natural disaster databases available, which primary interest is to assess insured losses.
While the number of events registered seems to confirm what described by our ground truth dataset (385 hydrological events in 2018, 340 in 2017), the geographical distribution of events over the globe shown in Figure~\ref{fig:munichre}  indicates a different distribution of events.
The discrepancies emerged can be explained in part by the fact that the majority of events comes from Floodlist, which also records  storms associated with heavy rainfall.
Such events fall into a different category of events (meteorological) in the NatCatSERVICE data.

One might think that our ground truth information could be biased by the coverage of events in wealthier countries where access to digital information is extensive and where English is the predominant language spoken.
Nevertheless, the EM-DAT distribution shown in Figure~\ref{fig:emdatmap} is similar to the one of NatCatSERVICE.

\subsection{Finding Floods in Wikipedia}

The next step was to try to locate these floods in Wikipedia.
Initially, a keyword-based search was done on articles from the English Wikipedia covering the period for which we collected ground truth information. To do this we used a public snapshot\footnote{\url{https://dumps.wikimedia.org} of the full English Wikipedia (containing around 6M articles), generated on May 20th, 2019.}.
We defined a set of keywords (``flood,'' ``floods,'' ``flooding,'' ``flooded,'' ``inundation'') as representative for a potential flood event identification in a sentence.
Therefore, we scanned for text containing any of the keywords within the sentences extracted from Wikipedia. If we found any of them, the sentence was stored with its corresponding title and paragraph as a ``candidate'' sentence.
In the case of articles whose title contained any of the keywords, such as “Floods in the United States: 2001–present”, all sentences were considered as candidates since we assumed that the content was about floods.

Of all the sentences filtered with the mentioned keywords, we selected only the ones directly linked to flooding. To ensure that, we applied to candidates a classifier (Lorini et al. 2019) expressing the probability of a text to be relevant to a flood.
We selected only the ones with a probability higher than 40\%. The Precision of this step was 83\%, computed over a sample of two hundred candidate sentences extracted from Wikipedia.

Selected articles needed to satisfy further criteria before being checked against ground truth information. Only the sentences containing information about Country and Time-span of the event(s) were considered potential candidates.
For extracting a date or location mentions, we used a Name Entity Recognition (NER) library named spaCy\footnote{\url{https://spacy.io}} on every title and sentence of the candidates.
Subsequently, all the potential candidates were parsed to extract timestamps and countries.
We created multiple candidates sentences in case spaCy returned multiple placenames.

For parsing string identified as dates, we used the datefinder\footnote{\url{https://datefinder.readthedocs.io}} library which can convert strings into structured objects.
It can also extract part of a date, such as ``early June''.
Since sometimes a specific year was not explicitly mentioned in a sentence but could be guessed from the context, we defined the following heuristics for extrapolating the year found elsewhere in a text:
\begin{itemize}[noitemsep,topsep=1pt]
  \item If there was only one year within a sentence, we could assume that the same year applies to every incomplete date entity in that sentence.
  \item If there was no year in the sentence and only one year in its whole paragraph, we could assume that the same year applies to that sentence.
  \item If there was no year in the sentence and only one year in the title of the article containing the sentence, it could be assumed that the same year applies to that sentence.
\end{itemize}
The heuristics used for associating place names to their respective country was a cascade of the following heuristics:
\begin{enumerate}[noitemsep,topsep=1pt]
\item Wikidata:
we searched for the placename identified by spaCy in Wikidata. If the entity returned has a corresponding page in the English Wikipedia, the country returned by the query is associated with the candidate sentence.
\item Nominatim:
we searched on Nominatim\footnote{\url{https://nominatim.openstreetmap.org/}} the place names that were not associated to a country after the first step.
The query used the public Nominatim API and the country associated with the place name was the most ``important''\footnote{The results have a value that represents the importance of the location according to the number of citations in Wikipedia.} result returned.
\item Mordecai:\footnote{\url{https://github.com/openeventdata/mordecai}}
the sentences and titles not associated with any country in the previous steps were then processed using Mordecai for inferring a country from the text.
\end{enumerate}
Tables~\ref{tab:countryparsing_f} shows the results of the application of our methodology on a set of sentences.
In some cases, the NER library could not find any placename; sometimes, the placename did not lead to the identification of the related country and in other cases, we could extract a country name.
Finally, we discarded candidate sentences for which we could not find a country and a time reference.

\begin{table}[htb]
\centering
\begin{tabular}{p{9cm}p{4cm}p{1.5cm}}
\toprule
 Sentence & Location entities & Countries \\ \midrule
``The 2009 West Africa floods are a natural disaster that began in June 2009 as a consequence of exceptionally heavy seasonal rainfall in large areas of West Africa'' & West Africa & None \\ \midrule
``In the Tiquicheo Municipality, 10 houses flooded after a river near the city overflowed its banks'' & the Tiquicheo Municipality & None \\ \midrule
``The town of Poldokhtar in Lorestan Province was engulfed by flood water.'' & Poldokhtar, Lorestan Province & None \\ \midrule
``2015 Southeast Africa floods'' & None & None \\ \midrule
``New Orleans Outfall Canals'' & None & None \\ \midrule
``Serious flooding was also reported in Greenwich, Woolwich and other locations further downriver, causing major property damage.'' & Greenwich, Woolwich & United Kingdom \\ \midrule
``In July 2012, heavy torrential rains caused floods in Kyushu, Japan, leaving 32 people dead or missing.'' & Kyushu, Japan & Japan \\ \midrule
``In Antu County, 70 homes in one village were destroyed by flooding, a mountain valley was submerged by floods 20 m deep, forcing 570 families to evacuate.'' & Antu County & China \\ \midrule
\end{tabular}
\caption{Examples of successful and unsuccessful attempts at location inference from text}
\label{tab:countryparsing_f}
\end{table}

\subsection{Matching Wikipedia Candidates and Ground Truth Information}
The last part of the matching process was determining if the selected Wikipedia sentences were identifying an entry in the ground truth database.
We defined three methods for identifying matching records. Here they are listed from the most strict to the laxest:

\begin{itemize}[noitemsep,topsep=1pt]
  \item Country and Year-Month-Days matching\\
  A Wikipedia candidate matches an event in the ground truth database if they link to the same country name and the date in the title or sentence of the candidate is within the time range[start\_date, end\_date+5 days] of the ground truth entry.

  For instance, the sentence ``On April 13, reportedly 12 people counted were killed by Rainstorm and Flash flooding in KPK and Balochistan.'' matches the flood happening in Pakistan between 2019-04-13 and 2019-04-18.
  \item Country and Month-Year matching\\
  A Wikipedia candidate matches an event in the ground truth database if they link to the same country name and the month in the candidate sentence or title is overlapping with the time range[start\_date, end\_date] of the ground truth entry.

  For instance, the sentence ``In August 2018, the region yet again experienced record-breaking flooding in valley towns such as Coon Valley, Wisconsin, La Farge, Wisconsin and Viola, Wisconsin.'' matches the flood happening between 2018-08-20 and 2019-08-22 in the USA.
\end{itemize}
After performing the matching for each of these pairing methods, we evaluated the hits manually for events covering three different months.

We define the \textbf{Precision} of our methodology as the fraction of matched candidates that are describing an event enlisted in the ground truth dataset.
We can think about Precision as the answer to the question \textit{How many Wikipedia matched candidates are a flood recorded in the ground truth dataset?}

We define the \textbf{Recall} of our methodology as the fraction of ground truth events that are identified by the matched candidates.
We can think about Recall as the answer to the question \textit{How many floods in the ground truth dataset are matched by Wikipedia candidates?}

Results of Precision and Recall evaluated manually over a sample of three months from our consolidated dataset are shown in Tables~\ref{tab:countrydate_ymd} and \ref{tab:countrydate_ym}.
\begin{table}[H]
\centering
\begin{tabular}{@{}llll@{}}
\toprule
 Period & Floods in ground truth & Precision(\%) & Recall (\%) \\ \midrule
 November 2016 & 18 & 66.67 & 16.67  \\
 September 2017 & 20 & 66.67 & 15.00 \\
 June 2018 & 26 & 88.89 & 34.62 \\
 \midrule
\end{tabular}
\caption{Metrics for country and year-month-days matching}
\label{tab:countrydate_ymd}
\end{table}
\begin{table}[H]
\centering
\begin{tabular}{@{}llll@{}}
\toprule
 Period & Floods in ground truth & Precision(\%) & Recall (\%) \\ \midrule
 November 2016 & 18 & 66.67 & 16.67  \\
 September 2017 & 20 & 50.00 & 20.00 \\
 June 2018 & 26 & 53.33 & 57.68 \\
 \midrule
\end{tabular}
\caption{Metrics for country and year-month matching}
\label{tab:countrydate_ym}
\end{table}
For the identification of a correlation between socio-economic indicators and flood coverage in Wikipedia, we opted for the matching method using Country and Year-Month-Days because a higher Precision implies that more matches are relevant, thus better support our analysis.

\section{Results}
\label{sec:results}
We will use the term \textbf{Hit Rate} to refer to the percentage of matches between the English Wikipedia articles and events registered in the ground truth database.
A Hit Rate of 100 means that applying the methodology presented in the previous section, all the floods representing a set of events in the ground truth database matched some Wikipedia candidates. A Hit Rate equal to 0 represents no coverage in the English Wikipedia for any flood of the set of events analyzed.

Our research analyzed how articles in English Wikipedia covered the floods reported worldwide in our ground truth database.
We analyzed several socio-economic variables to see whether they correlate with floods coverage.
These variables are GDP per capita, GNI per capita, country, continent, date, fatalities, number of English speakers and vulnerability index.

\begin{figure}[H]
\centering
  \includegraphics[width=.8\linewidth]{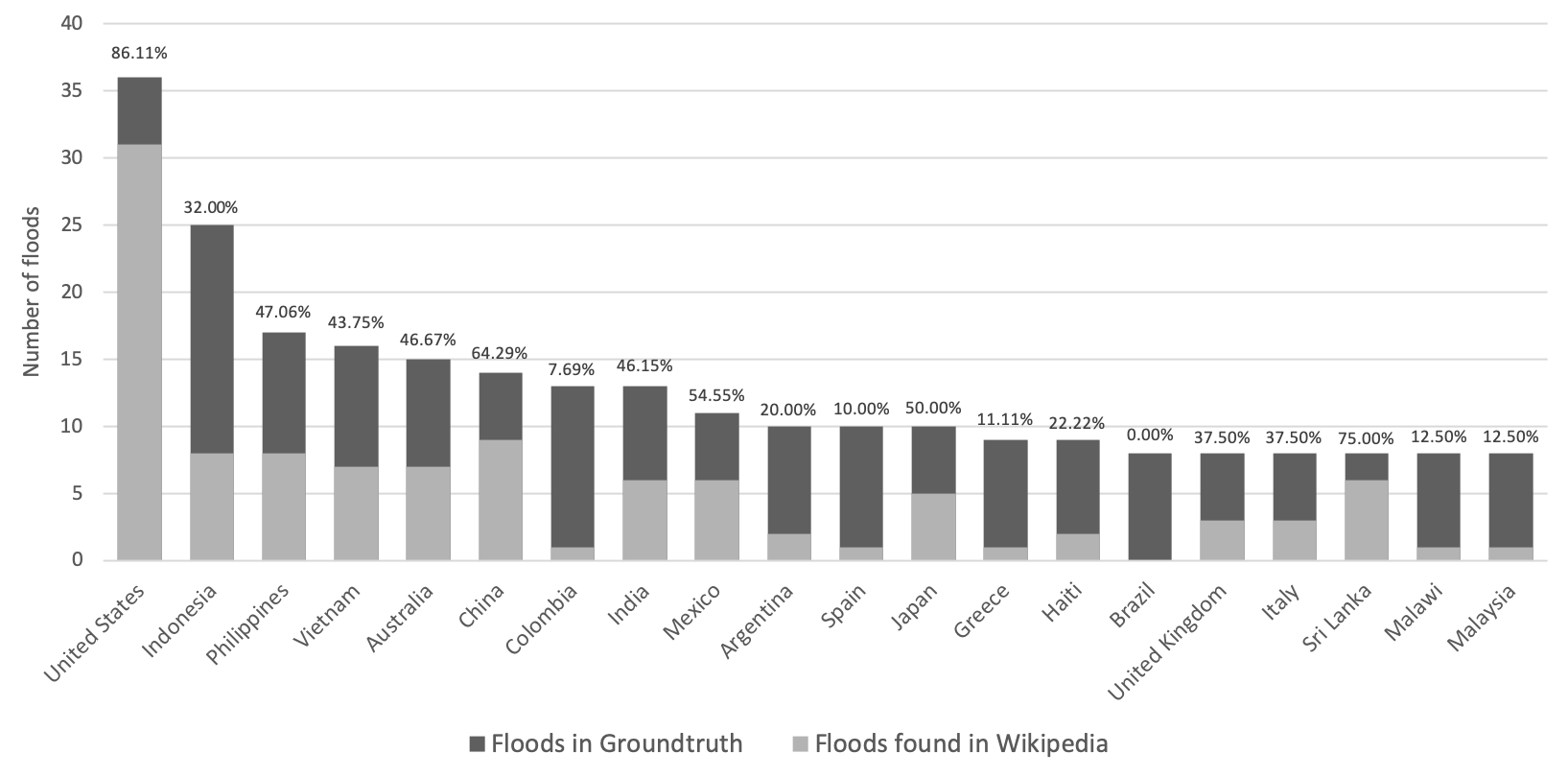}
  \caption{Top 20 countries ordered by number of floods in ground truth}
  \label{fig:countriesbygt}
\end{figure}

Figure~\ref{fig:countriesbygt} shows the top twenty countries ordered by the number of floods in the ground truth dataset and their respective hit rate.
Among the countries with the highest number of floods, the United States shows a hit-rate two times higher than the ones of all the others.

\begin{figure}[H]
\centering
  \includegraphics[width=.8\linewidth]{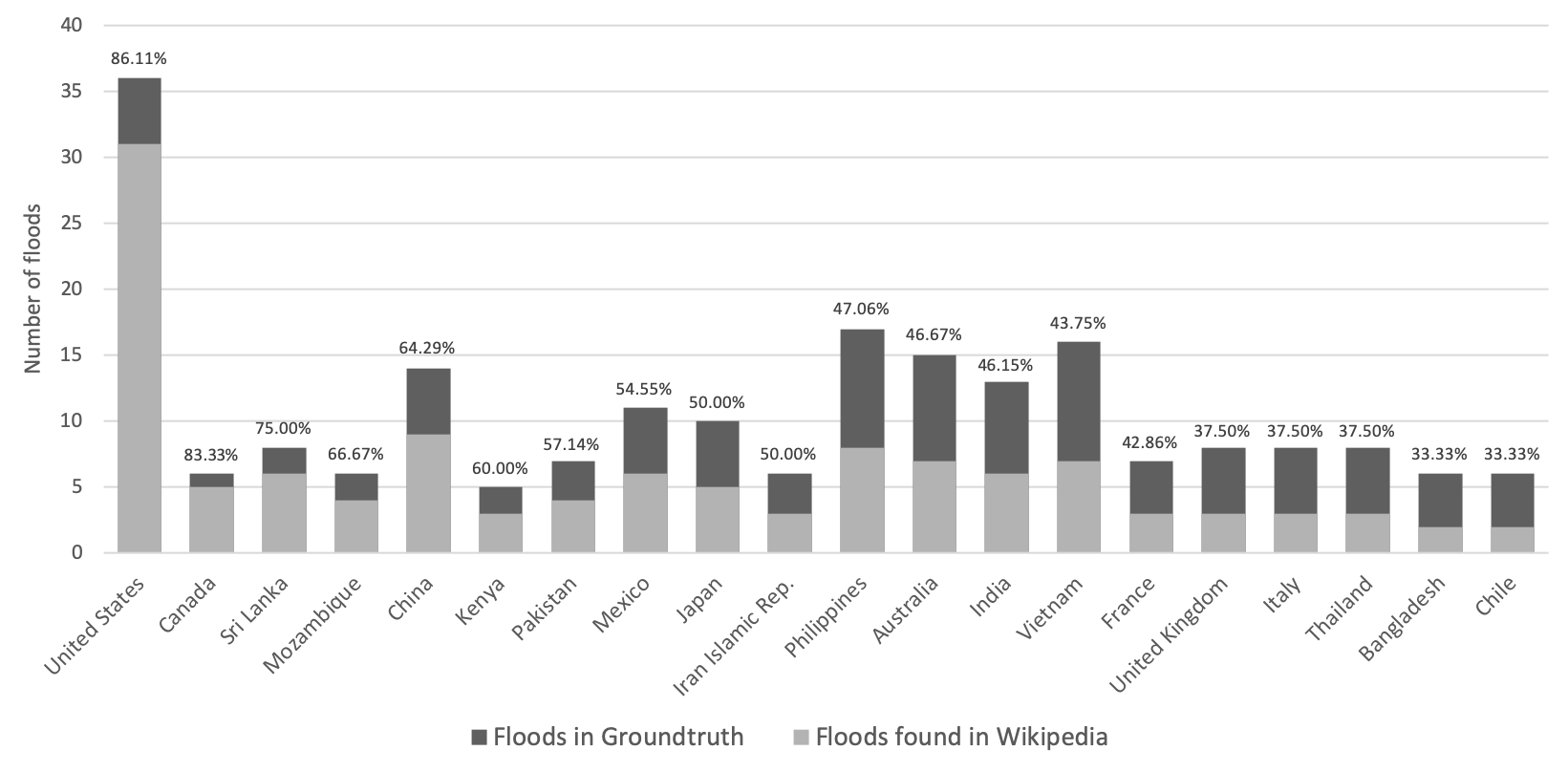}
  \caption{Top 20 countries with at least five floods in ground truth dataset ordered by hit rate}
  \label{fig:countriesbyhit}
\end{figure}

Figure~\ref{fig:countriesbyhit} shows the top twenty countries ordered by the hit-rate value and their respective coverage in the ground truth database. Although most of the floods in the ground truth database were detected in Wikipedia for more than half of them, in North American English-speaking countries the rate is above 80\%.
Floods events in the USA and Canada are reported most frequently on Wikipedia English than anywhere else in the world. The language can be only a partial explanation because for floods in Australia the hit-rate is half and lower than other non-English-speaking countries.

\begin{figure}[H]
\centering
  \includegraphics[width=.6\linewidth]{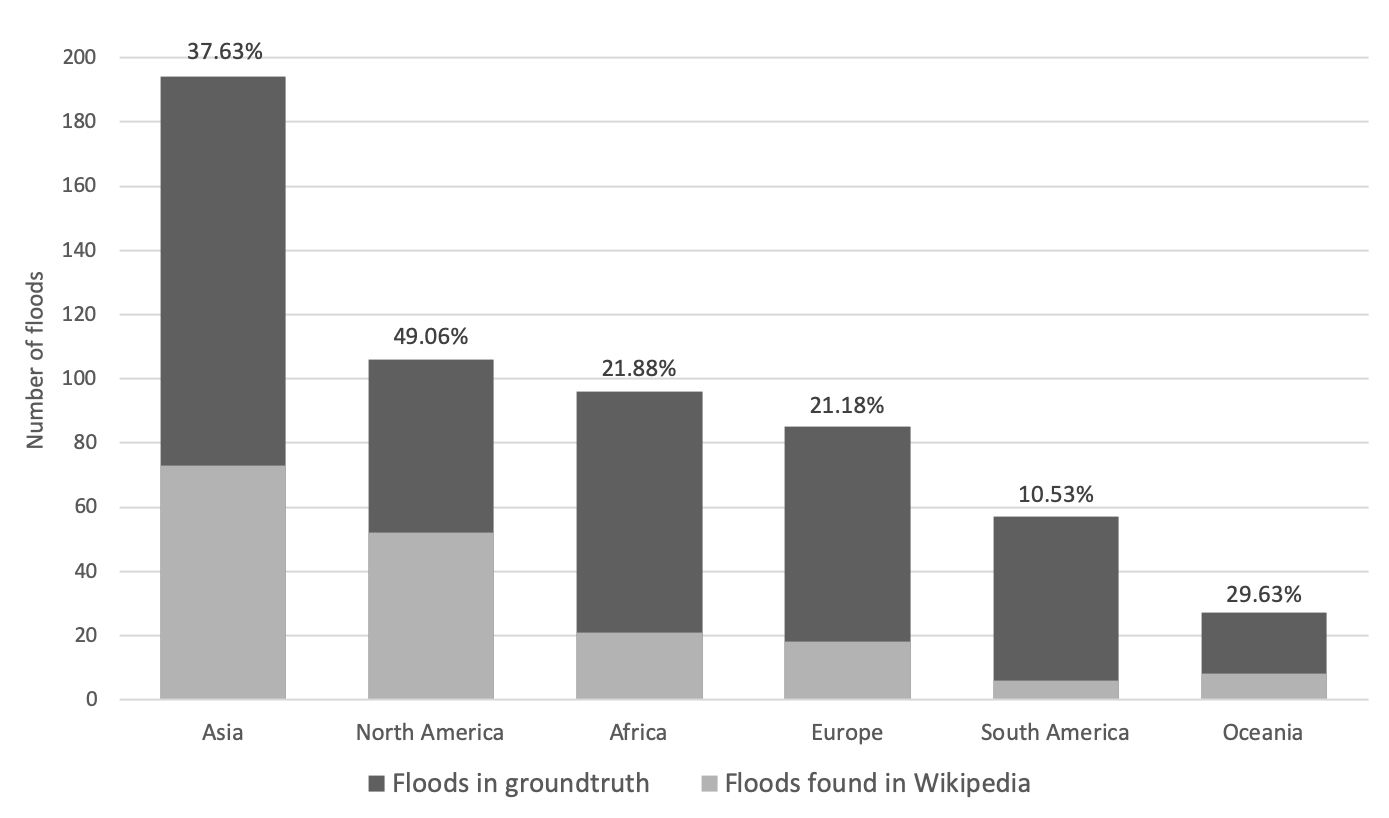}
  \caption{Floods for each continent and their corresponding hit rate ordered by number of floods in the ground truth}
  \label{fig:continentshit}
\end{figure}

Table~\ref{tab:continent} and Figure~\ref{fig:continentshit} show the ratio between floods recorded in the ground truth database and the floods detected in Wikipedia aggregated by continent. Floods are geophysical event. Thus this aggregation offer a comparison between similar Areas' extension.

Although most events happened in Asia, floods in North America have been reported more frequently.

\begin{table}[H]
\centering
\begin{tabular}{@{}llll@{}}
\toprule
 & Floods in ground truth & Floods in Wikipedia & Hit rate (\%)  \\ \midrule
 Asia & 194 & 73 & 37.63 \\
 North America & 106 & 52 & 49.06 \\
 Africa & 96 & 21 & 21.88 \\
 Europe & 85 & 18 & 21.18 \\
 South America & 57 & 6 & 10.53 \\
 Oceania & 27 & 8 & 29.63 \\ \bottomrule
\end{tabular}
\caption{Percentage of ground truth floods matched by Wikipedia: per Continent}
\label{tab:continent}
\end{table}

In order to deepening our analysis, we divided the countries into six different groups according to their Gross Domestic Product per capita in US Dollars, following the classification set by the World Bank for this indicator\footnote{url{https://data.worldbank.org/indicator/ny.gdp.pcap.cd}}:
\begin{itemize}[noitemsep,topsep=1pt]
  \item Low income: GDP per capita $<$ \$812
  \item Low middle income: \$812 $\le$ GDP per capita $<$ \$2,218
  \item Middle income: \$2,218 $\le$ GDP per capita $<$ \$5,484
  \item Upper middle income: \$5,484 $\le$ GDP per capita $<$ \$9,200
  \item High income: \$9,200 $\le$ GDP per capita $<$ \$44,714
  \item Very high income: GDP per capita $\ge$ \$44,714
\end{itemize}

\begin{figure}[H]
\centering
  \includegraphics[width=.7\linewidth]{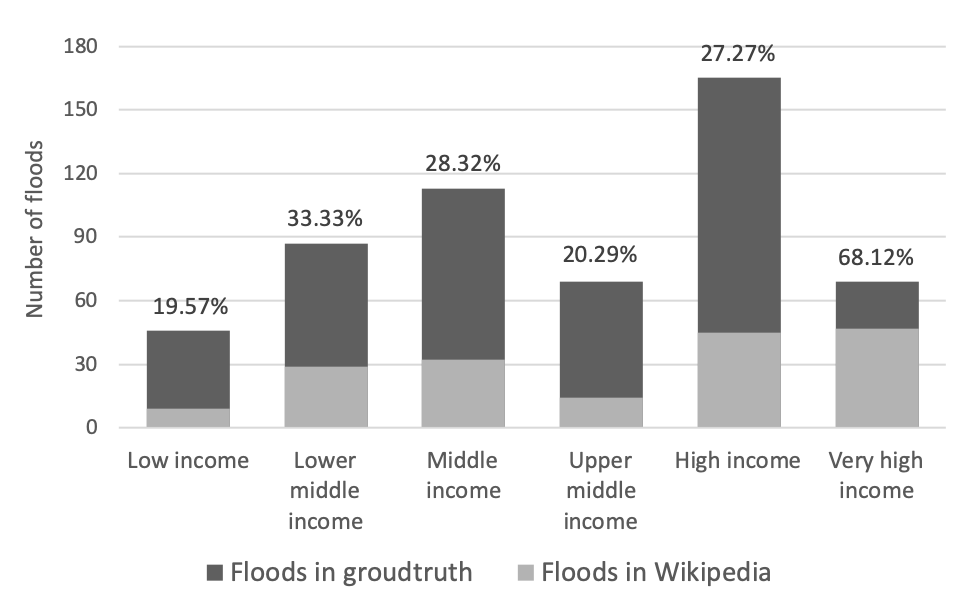}
  \caption{Floods for each level of GDP per capita and its corresponding hit rate}
  \label{fig:gdpcountryfig}
\end{figure}

Table~\ref{tab:gdpcountry} and Figure~\ref{fig:gdpcountryfig} show the results for each of these groups.
Within the ground truth database, floods recorded for countries in the high-income group are more than in any other group. The hit-rate of most countries from lower-middle-income to high-income varies between 30\% and 40\%.
Hit-rate is considerably different between the lowest and the highest bracket, close to 19\% for the former and 65\% for the latter.

\begin{figure}[H]
\centering
  \includegraphics[width=.5\linewidth]{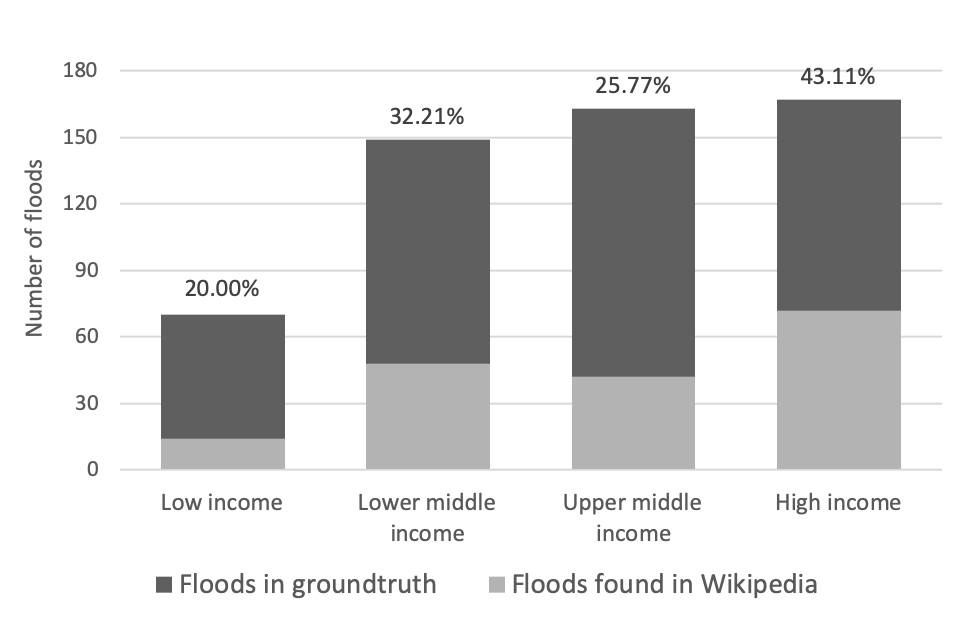}
  \caption{Floods for each level of GNI per capita and its corresponding hit rate}
  \label{fig:gnicountryfig}
\end{figure}

\begin{table}[H]
\centering
\begin{tabular}{@{}lllll@{}}
\toprule
 & Countries & Floods in ground truth & Floods in Wikipedia & Hit rate (\%)  \\ \midrule
 Low income & 21 & 46 & 9 & 19.57 \\
 Lower middle income & 37 & 87 & 29 & 33.33 \\
 Middle income & 39 & 113 & 32 & 28.32 \\
 Upper middle income & 27 & 69 & 14 & 20.29 \\
 High income & 67 & 165 & 45 & 27.27 \\
 Very high income & 28 & 69 & 47 & 68.12 \\ \bottomrule
\end{tabular}
\caption{Number of floods matched per Country according to GDP per capita}
\label{tab:gdpcountry}
\end{table}
We also grouped the countries into four different groups
following the classification set by the World Bank for this indicator\footnote{\url{https://data.worldbank.org/indicator/NY.GNP.PCAP.CD?view=chart}}.
The difference in coverage (hit rate) between high-income and low-income countries is even more evident, as shown in Figure~\ref{fig:gnicountryfig}.

INFORM Global Risk Indicators (GRI) is an open-source risk assessment tool for humanitarian crises and disasters. It can support decisions about prevention, preparedness and response \parencite{Ferrer2017}.
We extracted and combined two socio-economic indicators that can complement our previous analysis from their 2018 report:
\begin{itemize}[noitemsep,topsep=1pt]
\item The \textit{Vulnerability} indicator addresses the intrinsic predispositions of an exposed population to be affected or to be susceptible to the damaging effects of a hazard. So, the Vulnerability dimension represents the economic, political and social characteristics of the community that can be destabilized in case of a hazardous event.
Physical vulnerability is a different matter; it is embedded into the hazard and exposure indicators.
\item The \textit{Lack  of coping  capacity} indicator measures the ability of a country to cope with disasters in terms of formal, organized activities and the effort of the country’s government as well as the existing infrastructure, which contribute to the reduction of disaster risk.
\end{itemize}

They are both expressed on a scale of zero to ten. We combine the indicators as the square root of the product between them.
We grouped the events in the ground truth database into four categories sorted by ascending value of the combined indicator. The higher is the indicator, the more vulnerable is the country.%
\begin{table}[H]
\centering
\begin{tabular}{@{}llll@{}}
\toprule
 Vulnerability & Floods in ground truth & Floods in Wikipedia & Hit rate (\%)  \\ \midrule
 0-2 (least vulnerable) & 66 & 21 & 31.82 \\
 2-4 & 157 & 60 & 38.22 \\
 4-6 & 130 & 43 & 33.08 \\
 6-8 & 101 & 31 & 30.69 \\
 8-10 (most vulnerable) & 90 & 20 & 22.22 \\ \bottomrule
\end{tabular}
\caption{Percentage of ground truth floods matched by Wikipedia: per INFORM indicators}
\label{tab:copingvuln}
\end{table}
We can see how the most vulnerable group (8-10) is also the one with the lowest hit rate by a substantial margin.
Events where the capacity to cope with a disaster is the lowest, therefore where the impact could be the highest, are less likely to be described in Wikipedia.

\begin{figure}[H]
\centering
  \includegraphics[width=.5\linewidth]{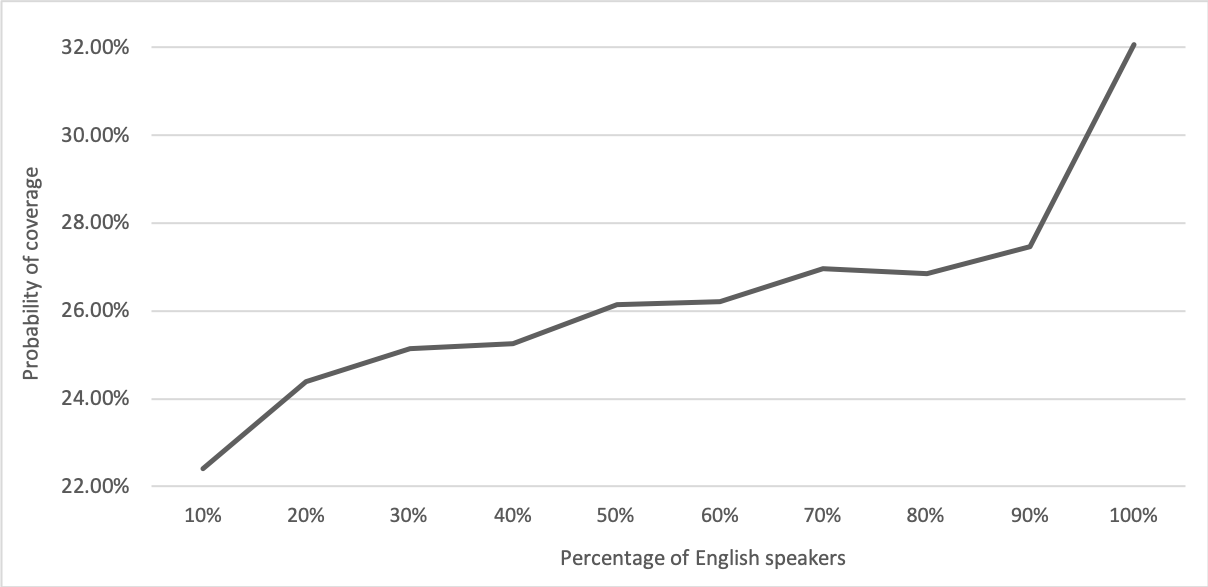}
  \caption{Probability of hit given the percentage of English speakers}
  \label{fig:speakersfig}
\end{figure}

\begin{figure}[H]
\centering
  \includegraphics[width=.7\linewidth]{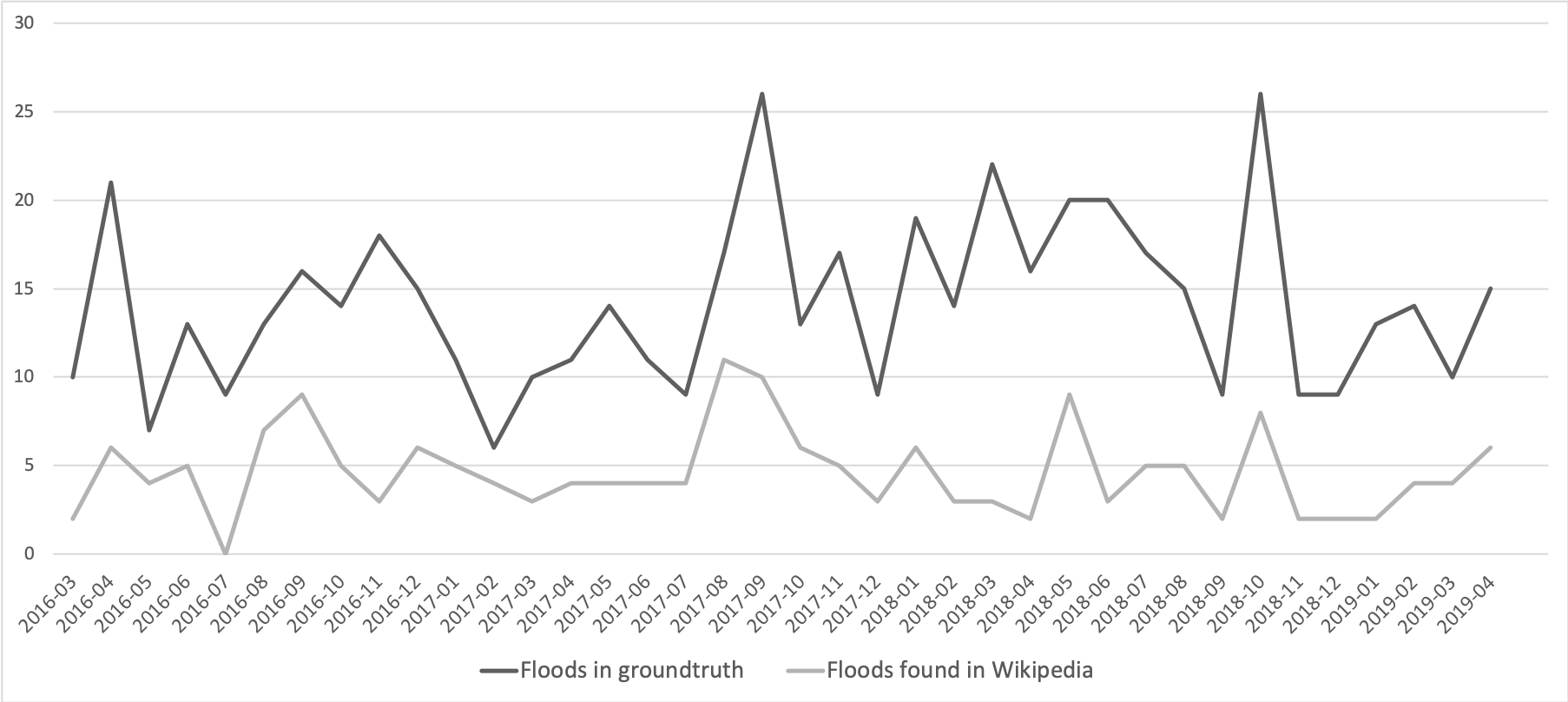}
  \caption{Floods for each month between 2016-03 and 2019-04}
  \label{fig:monthsfig}
\end{figure}

\begin{figure}[H]
\centering
  \includegraphics[width=.5\linewidth]{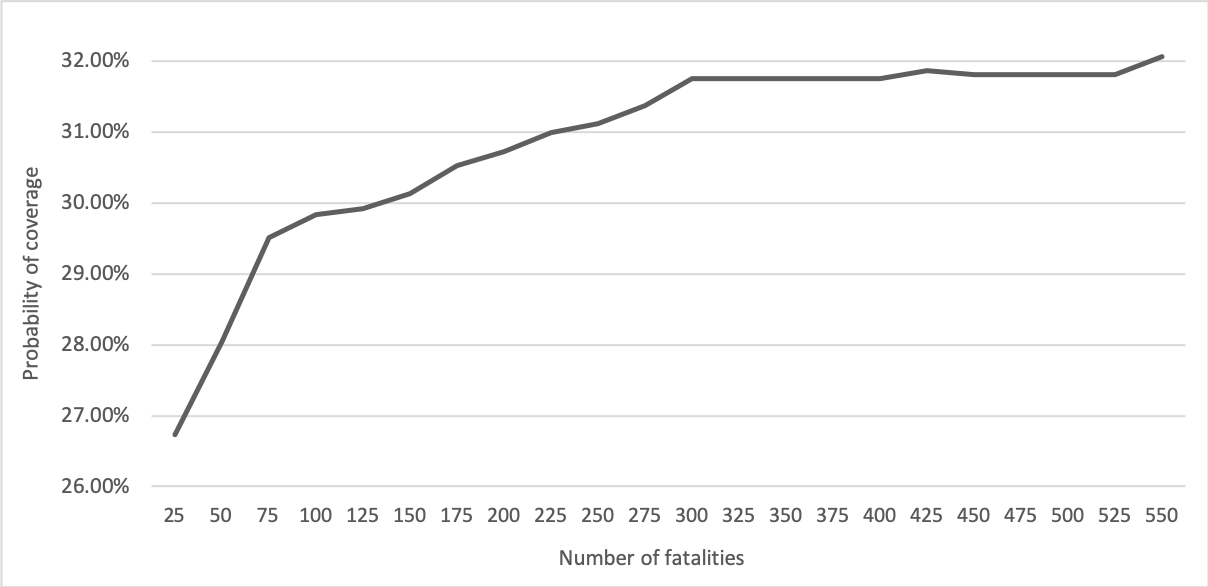}
  \caption{Probability of hit given the number of fatalities}
  \label{fig:fatalities}
\end{figure}

Our analysis also considered the percentage of English speakers\footnote{\url{ https://en.wikipedia.org/wiki/List_of_countries_by_English-speaking_population}} in a country.
We assumed that Floods in countries with a large percentage of English speakers were more likely to be matched in the English Wikipedia because of the availability of news in English that Wikipedia editors can understand and use as sources for documenting the floods.
Indeed, in Table~\ref{tab:enspeaker} and Figure~\ref{fig:speakersfig}, we see an increase in the coverage for countries with 60\% or more English speakers, and then another increase for countries with 80\% or more English speakers.
\begin{table}[H]
\centering
\begin{tabular}{@{}llll@{}}
\toprule
  English speakers (\%) & Floods in ground truth & Floods in Wikipedia & Hit rate (\%)  \\ \midrule
 <20 & 131 & 36 & 27.48 \\
 20-40 & 67 & 19 & 28.36 \\
 40-60 & 37 & 14 & 37.84 \\
 60-80 & 24 & 9 & 37.50 \\
 80+ & 92 & 53 & 57.61 \\ \bottomrule
\end{tabular}
\caption{Percentage of ground truth floods matched by Wikipedia: per percentage of English speakers}
\label{tab:enspeaker}
\end{table}
The percentage of the English-speaking population is an indicator of the probability that an event would be described in English Wikipedia.
Nevertheless, the population of a country could be related to the event coverage by English-speaking editors.
We sorted the countries by ascending population and divide them into four groups containing the same number of countries each.
\begin{itemize}[noitemsep,topsep=1pt]
  \item Group 1: population < 754,394
  \item Group 2: 754,394 $\le$ population < 6,465,513
  \item Group 3: 6,465,513 $\le$ population < 24,992,369
  \item Group 4: 24,992,369 $\le$ population
\end{itemize}
\begin{table}[H]
\centering
\begin{tabular}{@{}llll@{}}
\toprule
  Country population & Floods in ground truth & Floods in Wikipedia & Hit rate (\%)  \\ \midrule
 G1 (smallest) & 14 & 2 & 14.29 \\
 G2 & 60 & 8 & 13.33 \\
 G3 & 133 & 29 & 21.80 \\
 G4 (largest) & 342 & 137 & 40.06 \\ \bottomrule
\end{tabular}
\caption{Percentage of ground truth floods matched by Wikipedia: per population}
\label{tab:populationhit}
\end{table}
Indeed, as shown in Table~\ref{tab:populationhit}, the hit rate is more significant for the most populated countries.
In order to determine if the time of events affects the coverage of floods in Wikipedia, we analyzed the temporal distribution of the events.
The method applied for evaluating hit rates showed that the relation between ground truth events and matches follow similar proportions across time, as shown in Figure~\ref{fig:monthsfig}.
Table\ref{tab:fatalities} and Figure~\ref{fig:fatalities} shows a significant increase in the number of events that are matched by Wikipedia articles for floods in the ground truth database leading to  hundred of fatalities or more.
Combined with the other indicators, this could mean that only events with high impacts echoed to countries with higher English-speaking population rate and high-income to mid-income.
\begin{table}[H]
\centering
\begin{tabular}{@{}llll@{}}
\toprule
 Num. of fatalities & Floods in ground truth & Floods in Wikipedia & Hit rate (\%)  \\ \midrule
 0 & 164 & 31 & 18.90 \\
 1-9 & 184 & 57 & 30.98 \\
 10-99 & 173 & 68 & 39.31 \\
 100-1999 & 27 & 20 & 74.07 \\ \bottomrule
\end{tabular}
\caption{Percentage of ground truth floods matched by Wikipedia:per number of fatalities}
\label{tab:fatalities}
\end{table}
\label{sec:domains}
Finally, since Wikipedia editors must reference external sources for citations (according to Wikipedia guidelines), we decided to explore from where they retrieve their information.
We extracted all domains included as references in any sentence matching a flood in the ground truth dataset. The top ten sources used by Wikipedia editors are listed in Table~\ref{tab:domainstab}.
\begin{table}[H]
\centering
\begin{tabular}{@{}llll@{}}
\toprule
 Domain &  References \\ \midrule
 indiatimes.com & 20 \\
 weather.com & 19 \\
reliefweb.int & 15 \\
abc.net.au	 & 15 \\
theguardian.com & 14 \\
reuters.com & 14 \\
bbc.com & 13 \\
hindustantimes.com & 13 \\
cnn.com & 12 \\
dailymirror.lk & 12 \\ \bottomrule
\end{tabular}
\caption{Top ten sources from which Wikipedians retrieve information about floods}
\label{tab:domainstab}
\end{table}
It was surprising that the most mentioned sources from the ones we used, Floodlist, did not appear among the top ten and was only referenced four times in our dataset. Besides \emph{reliefweb.int}, we do not see in the top-10 list specialized sources on disaster management. This can be explained by the fact that our sources record events only when data are validated by an official source, which usually happens after the peak of the event has ended.
\section{A Tale of two Floods} %
\spara{North and Central America, September 2017.}
Hurricane Irma made landfall on northeast Caribbean islands during the early hours of 6 September, affecting Antigua and Barbuda, Anguilla, British Virgin Islands, St Barthélemy, St. Martin, the Virgin Islands and other islands in the eastern Caribbean Sea.
After causing devastating damage across the Caribbean, Hurricane Irma made landfall in the Florida Keys on 10 September and worked its way north, bringing with it strong winds, storm surge and flooding rain.

Although national news agencies covered only partially the event, we can say that Irma caused between fifty and one-hundred fatalities, affecting millions of people (evacuated or left without power).
In our ground truth database, the Hurricane Irma event is linked to several countries, and it was one of the events that produced the highest number of matches (43 total, 40 only USA), meaning many sentences on Wikipedia reported about it. Find some examples of sentences reporting Hurricane Irma in Table \ref{tab:irmaexamples}

\begin{table}[H]
\centering
\begin{tabular}{p{10cm}p{2cm}p{2cm}}
\toprule
 Sentence & Date & Countries \\ \midrule
``In September 2017, Hurricane Irma storm surge caused major flooding in the downtown area of Jacksonville.'' & 2017-09-12 & United States of America \\ \midrule
``Moyer, Crystal (September 8, 2017). Hyatt Regency in downtown Jacksonville being evacuated.'' & 2017-09-08 & United States of America \\ \midrule
``People stand in a flooded street that usually serves as a farmers market, in Ouanaminthe, northeast Haiti, September 8, 2017.'' & 2017-09-08 & Haiti \\ \midrule
``Hurricane Irma: 10 dead in Cuba as record flooding hits northern Florida [...] September 11, 2017.'' & 2017-09-11 & Cuba \\ \midrule
\end{tabular}
\caption{Examples of sentences reporting Hurricane Irma}
\label{tab:irmaexamples}
\end{table}

\spara{Sudan, August 2018.}
By August 2018 heavy rains in Sudan that had started in mid-July had caused severe flooding. As of 16 of August the floods and rain had left at least 23 people dead, over 60 injured and affected more than 70,000 people.
Although the event appears in all three data sources, we could not find any match while in Wikipedia applying our methodology.  Either the event was not the subject of any Wikipedia article, or it was not described as accurately as other events.

Even if both events had a high number of fatalities and affected people, while the former event was widely identifiable on Wikipedia, the latter case was less (or poorly) described.

\section{Conclusions and Future Work} %

According to the United Nations Office for Disaster Risk Reduction, the impact of natural hazards is highest on the most marginalized populations, exacerbating inequality and further entrenching poverty, where vulnerabilities translate into reduced access to entitlements, impaired capabilities and opportunities.
For instance, it is estimated that 35.6\% of the population affected by floods in Pakistan in 2010 consequently slipped under the poverty line as a result.
Beyond focused attribution to single events, impacts are often found to be a function of a series of associated shocks - famine, disease and displacement for instance - that collectively prompt disruption in multiple dimensions (e.g. livelihoods, educational trajectories or labour-market opportunities)
\parencite{UNDRR_2019}.

The results of our analysis are consistent along several dimensions, and paint a picture in which Wikipedia's coverage is biased towards some countries, particularly those that are more industrialized and have large English-speaking populations, and against some countries, particularly low-income countries which also happen to be among the most vulnerable.
This points out that tools using data from social media or collaborative platforms should be carefully evaluated for biases, and that Wikipedia editors need to put more effort in covering disasters in the countries that most need it.

\spara{Limitations and future work.}
We considered only one type of event that is very prevalent globally: floods, but other types of events should be considered.
We had chosen to focus on one type of event because this work is a first attempt to bring Wikipedia and its crowdsourced opensource information into the scope of Disaster Management. Therefore the experiments and the results must be solid, reproducible and clear.

We relied on methodologies demonstrated and developed in previous work such as automated classification of text using ML models and we used consistent exhaustive data sources.
We strongly believe Wikipedia would be beneficial for researcher and crisis responders eventually, that is why we assessed a series of issues to be addressed and measured in order to pave the road for its fruitful use in Crisis rooms.

Automated content analysis cannot replace expert annotation, but considering that the English Wikipedia contains over five million articles, it is impractical to perform this analysis manually.
Some biases introduced by automated content analysis may include the usage of libraries for parsing geographical entities, which may have been trained using more data from some countries than from others; these biases need to be studied.

It would be necessary to perform this study considering other (language) versions of Wikipedia in order to understand how an editor's language affects the coverage bias.
The scope of the research must be extended to other types of events such as forest fires. In order to scale the current methodology, researchers need to identify an adequate global ground truth information  for the specific type of event and an automated text classifier for it.
An additional research effort could analyze the information derived from  events that are reported in the ground truth database but are not found in Wikipedia.

\spara{Reproducibility.}
Code implementing our methods, the merged list of floods, and the raw and processed datasets of Wikipedia matches are available: \url{https://github.com/javirandor/disasters-wikipedia-floods}

\spara{Acknowledgments.}
Castillo thanks La Caixa project LCF/PR/PR16/11110009 for partial support.
Rando was partially funded by the María de Maeztu Units of Excellence Program (MDM-2015-0502).

\printbibliography

\end{document}